\documentclass[9pt,journal]{IEEEtran}
\usepackage{float}
\usepackage[table]{xcolor}
\usepackage{cite}      
\usepackage{graphicx}  
\usepackage{psfrag}    
\usepackage{hyperref}
\usepackage{cleveref}
\usepackage{placeins}
\usepackage{url}
\usepackage{stfloats}
\usepackage{subfigure}
\usepackage{multirow}
\usepackage{caption,tabularx,booktabs}
\usepackage{multicol}
\DeclareRobustCommand{\bigO}{%
  \text{\usefont{OMS}{cmsy}{m}{n}O}%
}
\usepackage{times}
\usepackage{textcomp}
\usepackage{amsmath}   
\usepackage{amsfonts,amsthm,amssymb}
\usepackage{mathtools}
\usepackage{nicefrac}
\usepackage[none]{hyphenat}
\usepackage{balance}
\usepackage{cite}
\usepackage{array}
\usepackage{fontenc}
\usepackage{color}
\usepackage{etoolbox}
\usepackage{bm}
\begin{document}
\title{Mitigation of Through-Wall Distortions of Frontal Radar Images using Denoising Autoencoders}
\author{Shelly~Vishwakarma,~\IEEEmembership{Student Member,~IEEE,}
    Shobha~Sundar~Ram,~\IEEEmembership{Member,~IEEE}
\thanks{The authors are with Indraprastha Institute of Information Technology Delhi (email: shellyv@iiitd.ac.in; shobha@iiitd.ac.in).}}

\maketitle

\begin{abstract}
Radar images of humans and other concealed objects are considerably distorted by attenuation, refraction and multipath clutter in indoor through-wall environments. While several methods have been proposed for removing target independent static and dynamic clutter, there still remain considerable challenges in mitigating target dependent clutter especially when the knowledge of the exact propagation characteristics or analytical framework is unavailable. In this work we focus on mitigating wall effects using a machine learning based solution- denoising autoencoders- that does not require prior information of the wall parameters or room geometry. Instead, the method relies on the availability of a large volume of training radar images gathered in through-wall conditions and the corresponding clean images captured in line-of-sight conditions. During the training phase, the autoencoder learns how to denoise the corrupted through-wall images in order to resemble the free space images. We have validated the performance of the proposed solution for both static and dynamic human subjects. The frontal radar images of static targets are obtained by processing wideband planar array measurement data with two-dimensional array and range processing. The frontal radar images of dynamic targets are simulated using narrowband planar array data processed with two-dimensional array and Doppler processing. In both simulation and measurement processes, we incorporate considerable diversity in the target and propagation conditions. Our experimental results, from both simulation and measurement data, show that the denoised images are considerably more similar to the free-space images when compared to the original through-wall images.
\end{abstract}
\providecommand{\keywords}[1]{\textbf{\emph{Keywords--}}#1}

\IEEEpeerreviewmaketitle

\section{Introduction}
\label{Sec:Intro}
Through-the-wall radar imaging (TWRI) has been extensively researched in recent years, for detecting and monitoring humans and other concealed objects in urban environments. There are varied applications for TWRI such as law enforcement, security, and surveillance, search and rescue, and indoor monitoring of the elderly \cite{amin2016through,narayanan2010through,ram2008through, ssramthroughwall,lai2008hilbert}. 
There are broadly two categories of through-the-wall radars: narrowband and broadband. Broadband radars provide excellent downrange resolutions to locate and resolve multiple targets as well as for estimating building layouts \cite{jia2014multichannel}. Alternately, narrow band continuous wave (CW) radars have been developed to detect dynamic targets based on their Doppler signatures \cite{clemente2013developments,narayanan2010through,ram2008through}. Both of these systems can be complemented with two-dimensional array processing to provide either range-enhanced frontal images or Doppler-enhanced frontal images \cite{ram2015high,ram2016through}. Frontal images of the humans provide informative signatures of their activities \cite{ram2015doppler}. However, when the radars are deployed in through-wall scenarios, the quality of the radar images significantly deteriorate due to the through-wall propagation artifacts such as - attenuation, defocussing and multipath clutter \cite{ram2008through,ram2015high,ahmad2007autofocusing,leigsnering2014multipath}. 

Indoor clutter can be broadly categorized into \emph{target independent static and dynamic clutter}, and \emph{target dependent clutter}.
\emph{Target independent static clutter} arise from the reflections off the wall (especially the front face in a through-wall scenario), ceiling, floor, and furniture. Static clutter is easy to eliminate through filtering when the objective is to detect dynamic targets. The problem becomes more challenging in the context of detection of static and slow-moving targets. Authors in \cite{ahmad2008multi} assumed the availability of background data that could be coherently subtracted from the target measurements.
Alternately, sparsity-based multipath exploitation methods were explored in \cite{leigsnering2018sparsity}. Here, the algorithm leveraged the orthogonality between the static clutter and the target scattering to mitigate the clutter.
\emph{Target independent dynamic clutter} arising from other moving objects in the environment can significantly interfere with Doppler signatures of moving targets. In \cite{shellytrans2016}, a method to segregate the Doppler returns from multiple targets was presented. This technique could be used for mitigating dynamic clutter. The third category is \emph{target dependent clutter} that arises from the interactions of the target (static or dynamic) and the complex propagation channel. As a result of refraction and multipath, the radar images are smeared, blurred, and there are  shifts in the location of point scatterers in the images \cite{ram2015high}. The authors in \cite{martone2009through} and \cite{ahmad2013through} used  back-projection and sparsity based change detection algorithms, respectively, to  track slow moving humans in the range-crossrange space in the presence of target dependent clutter. Both these techniques, however, rely on the availability of accurate knowledge of the through-wall scenarios for detecting static targets. 
Alternately, the authors in \cite{setlur2011multipath,setlur2013multipath} exploited the multipath (instead of suppressing the multipath) to improve the the effective signal-to-clutter ratio (SCR) at the original target locations. They removed ghost artifacts by mapping the multipath ghosts to their true targets. Again the technique requires exact information of the room geometry and wall characteristics.

In this paper, we propose an alternate strategy, based on denoising autoencoders, for recovering radar images corrupted by through-wall effects.  An autoencoder is a neural network that extracts relevant features from the noisy input data for varied tasks such as- dimensionality reduction or data denoising \cite{vincent2008extracting,hinton2006reducing}. Autoencoders have been widely used for applications such as anomaly detection, natural language processing, denoising and domain adaptation\cite{sakurada2014anomaly,socher2011dynamic,vincent2008extracting,chen2012marginalized}. Some preliminary results for clutter mitigation using autoencoders were presented in  \cite{vishwakarma2018mitigation} where the nature of the type of through-wall scenario was assumed to be known during the test phase.
The primary advantage of this technique is, however, that the autoencoders require neither prior information regarding the wall characteristics nor any kind of analytic framework to describe the through-wall interference. Instead, the distorted radar signatures due to wall interference are treated as corrupt versions of ideal radar signatures obtained in free space conditions. The algorithm \emph{learns} how to denoise or clean the corrupted signals using training data comprising of both corrupted and clean data. We demonstrate, in this paper, that the autoencoder can be used for removing signal dependent clutter when no information or label of the through-wall scenario is assumed to be known during the test phase. Instead, the autoencoder is trained with a mixture of images gathered in diverse through-wall conditions.
Traditional autoencoders have been implemented using back-propagation algorithms such as- gradient descent \cite{du2017stacked}, conjugate gradient descent \cite{sutskever2013importance} and steepest descent \cite{hinton2006reducing}. However, they have a very slow learning rate. This translates to long training times and, in some cases, the convergence may not be guaranteed. Instead, we propose to use an alternating direction method of multipliers (ADMM) approach \cite{boyd2011distributed}, where we break the complex convex optimization problem into smaller sub-problems with closed-form solutions. Thus the convergence is guaranteed and training times are not very long.

We test the performance of the proposed algorithm on two types of radar images  - Doppler enhanced and range enhanced frontal images. The \emph{Doppler-enhanced frontal} images of \emph{dynamic human motions} are generated from simulated narrowband radar data of human motions in through-wall environments using the techniques described in \cite{ram2015high,ram2016through}. We consider a variety of walls-  a dielectric wall, a dielectric wall with metal reinforcements and one with air-gaps. The through-wall propagation phenomenology is modeled using finite difference time-domain (FDTD) techniques \cite{yee1966numerical}. We introduce significant diversity in wall parameters such as dielectric constant and conductivity by incorporating stochasticity in the finite difference equations as suggested by \cite{smith2012stochastic}. This is a computationally more efficient technique than running multiple FDTD simulations with varying wall parameters.  
The second set of images are  \emph{range-enhanced frontal images} captured of \emph{static humans} using measurement data gathered with Walabot, a three dimensional programmable, wideband imaging radar \cite{walabot_2017}.  
During the training phase, the autoencoder is trained with a diverse mixture of data gathered from different through-wall scenarios. In the test phase, the network denoises the corrupted radar image without requiring any information of the type of wall or its parameters. 
Both the simulation and measurement results obtained from the autoencoder exhibit very low normalized mean square error and high structural similarity between the denoised reconstructed images and free space images. 

To summarize, our contributions in this paper are the following: 
\begin{itemize}
\item First, we propose a denoising autoencoder to mitigate clutter and distortion in through-wall frontal images of both static and dynamic humans.
    \item Second, we propose a method to implement the autoencoder using ADMM approach to ensure convergence and fast training times.
\end{itemize}
In Section. II, we briefly describe the denoising autoencoder structure implemented in our work. Then, we use a computationally efficient method based on stochastic FDTD to simulate narrowband Doppler enhanced frontal images of dynamic humans as described in Section. III. Finally, we denoise range-enhanced through-wall radar images of static humans captured by a wideband RF sensor, Walabot, in Section. IV. We present the results, analyses and discussion on the strengths and limitations of the proposed approach in the final section. 
\section{Theory}
\label{Sec:Theory}
Radar images deteriorate significantly due to distortions and multipath clutter signals introduced by through-wall environments. The images may be defocused, blurred or smeared. Ghost targets may appear due to multipath. The objective, here, is to reconstruct clean radar images resembling free space images from corrupted through-wall images. We divide our denoising problem into two stages- training and the test stages. 
\begin{figure}[t]
\centering
\includegraphics[scale=0.6]{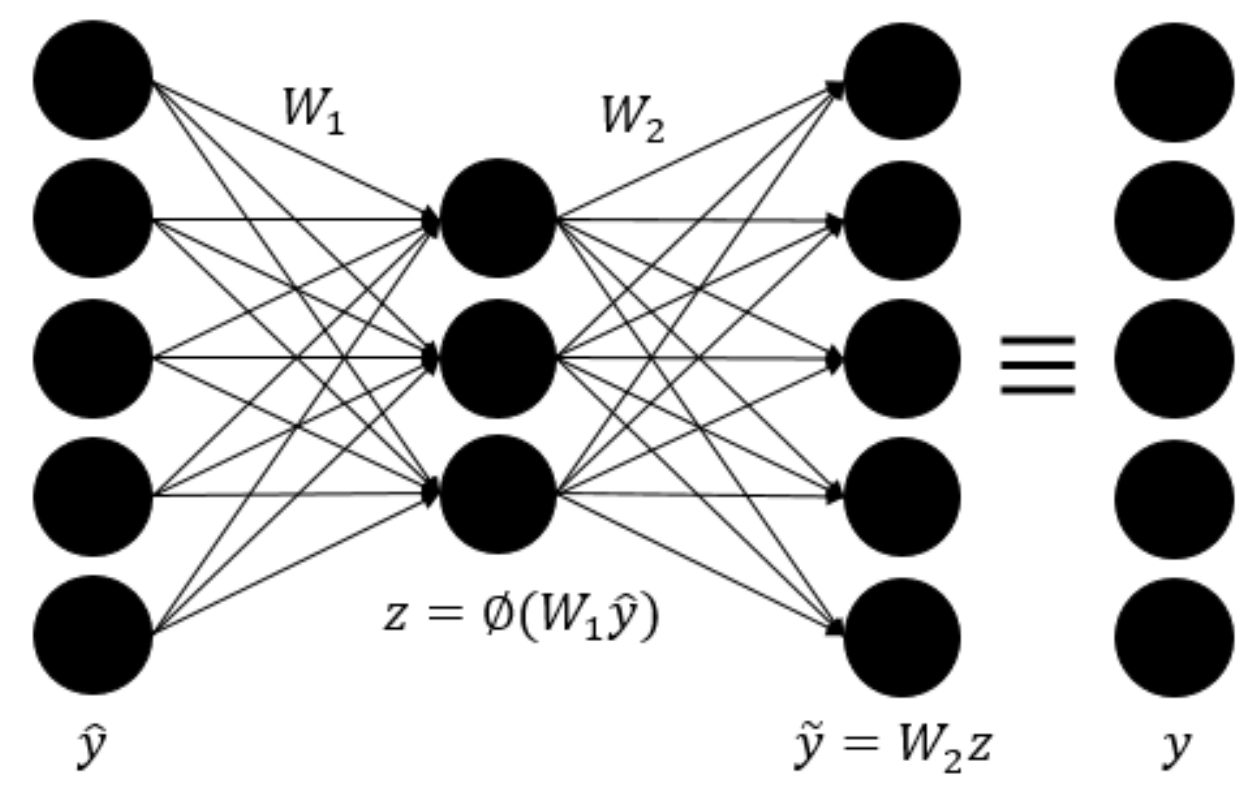}
\caption{Denoising Autoencoder}
\label{fig:denoising_autoencoder}
\end{figure}
 \subsection{Training Stage}
A conventional denoising autoencoder shown in Fig.\ref{fig:denoising_autoencoder}, first corrupts the clean input data by adding stochastic Gaussian noise, then feeds the corresponding noisy version as input data to the next stage. In this work, we consider the measurements in a through-wall case as our noisy/corrupted data. The main difference is the non-Gaussian nature of interference. 
The radar images captured in free space are $Y^{tr}\in \Re ^{N\times M}$ while the through-wall images are $\hat{Y}^{tr}$. Here, $M$ is the number of $N$-pixel radar images in both free space and through-wall scenarios. 
The autoencoder has primarily two stages- encoding and the decoding. In the encoding stage, the algorithm learns a latent/compressed representation $Z\in \Re ^{r\times M}$, of the input layer $\hat{Y}^{tr}$ as shown in (\ref{eqn:encode}).   
\begin{equation}
\label{eqn:encode} 
    Z=\phi(W_{1}\hat{Y}^{tr})
\end{equation}
Here, $\phi$ is the mapping function which can either be linear or nonlinear (such as sigmoid, $\tanh$), $W_{1}\in \Re ^{r\times N}$ the corresponding weight matrix and $r$ is the number of nodes in the hidden layer. Since the hidden layer, $Z$, is the compressed representation of the input layer, $\hat{Y}^{tr}$, it always has fewer nodes than the number of pixels ($r<<N$). In the decoding stage, the algorithm maps $Z$ back to obtain a reconstructed signal $\tilde{Y}^{tr}=W_{2}\phi(W_{1}\hat{Y})$ through weight matrix $W_{2}\in \Re ^{N\times r}$ such that the error $e$ is minimized in (\ref{eqn:error}). Note that the error is between the reconstructed images and the free space images.
\begin{equation}
\label{eqn:error} 
    e=\left \| Y^{tr}-\tilde{Y}^{tr}\right\|_{2}^2
\end{equation}
Therefore the objective in the training stage is to learn weight matrices $W_{1}$ and $W_{2}$ so that the reconstructed images resemble free space images (instead of the corrupted through-wall images). 
\begin{equation}
\label{eqn:ObjFunc} 
    J(W_1,W_2)=\min_{W_1,W_2} \left \|  Y^{tr}-W_2\phi(W_1\hat{Y}^{tr})\right \|_{2}^{2}
\end{equation}
The objective function (\ref{eqn:ObjFunc}) can be solved in multiple ways - gradient descent, conjugate gradient descent, or steepest descent. In some of these ways, the error may become insignificant when back propagated. Additionally, these algorithms have a very slow learning rate. Instead, we propose an alternating direction method of multipliers (ADMM) approach \cite{boyd2011distributed}. Here, we introduce a simple variable separation technique to break the complex convex optimization problem into smaller sub-problems which have closed form solutions such that the convergence is guaranteed. 
The objective function in (\ref{eqn:ObjFunc}) is reformulated to (\ref{eqn:ObjFunc1}).
\begin{eqnarray}
\begin{array} {lcl}
\label{eqn:ObjFunc1} 
    J(W_1,W_2)=&\min_{W_1,W_2} \left \|  Y^{tr}-W_{2}Z\right \|_{2}^{2}\\
    &s.t. \hspace{0.1cm} Z=\phi(W_1\hat{Y}^{tr})
\end{array}
\end{eqnarray}
Since the formulation in (\ref{eqn:ObjFunc1}) is a constrained optimization problem, we relax it using an augmented Lagrangian technique shown in (\ref{eqn:ObjFunc2}). 
\begin{eqnarray}
\begin{array} {lcl}
\label{eqn:ObjFunc2} 
    J(W_1,W_2,Z)=&\min_{W_1,W_2,Z} \left \|  Y^{tr}-W_2Z\right \|_{2}^{2}\\
    &+ \lambda\left \| Z-\phi(W_1\hat{Y}^{tr})\right \|_{2}^{2}
\end{array}
\end{eqnarray}
Here, $\lambda$ is the regularization parameter between the proxy variable $Z$ and underlying representation $\phi(W_1\hat{Y}^{tr})$. We divide (\ref{eqn:ObjFunc2}) into a set of smaller sub problems as follows.\\
Problem1:
\begin{equation}
\label{eqn:ObjFunc2a} 
   J(W_1)= \min_{W_1}\left \| \phi^{-1}Z-W_1\hat{Y}^{tr}\right \|_{2}^{2}
\end{equation}
Problem2:
\begin{equation}
\label{eqn:ObjFunc2b} 
    J(W_2)=\min_{W_2}\left \|  Y^{tr}-W_2Z\right \|_{2}^{2}
\end{equation}
Problem3:
\begin{eqnarray}
\begin{array} {lcl}
\label{eqn:ObjFunc2c} 
  J(Z)= &\min_{Z} \left \|  Y^{tr}-W_2Z\right \|_{2}^{2}+ \lambda\left \| Z-\phi(W_1\hat{Y}^{tr})\right \|_{2}^{2}\\
   &=\min_{Z}\begin{Vmatrix}
\begin{pmatrix}
Y^{tr}\\ 
\sqrt{\lambda}\phi(W_1\hat{Y}^{tr})
\end{pmatrix}-\begin{pmatrix}
W_2\\ 
\sqrt{\lambda}I
\end{pmatrix}{Z}
\end{Vmatrix}^{2}_{2}
\end{array}
\end{eqnarray}
Sub problems in (\ref{eqn:ObjFunc2a})-(\ref{eqn:ObjFunc2c}) are all simple least squares problems which already have a closed form solution \cite{bishop2006pattern}. At each iteration, we update the network weight $W_1$, $W_2$ and proxy variable $Z$, till the algorithm converges. 
\subsection{Test Stage}
\begin{figure}[t]
\centering \subfigure[]{
\includegraphics[scale=0.5]{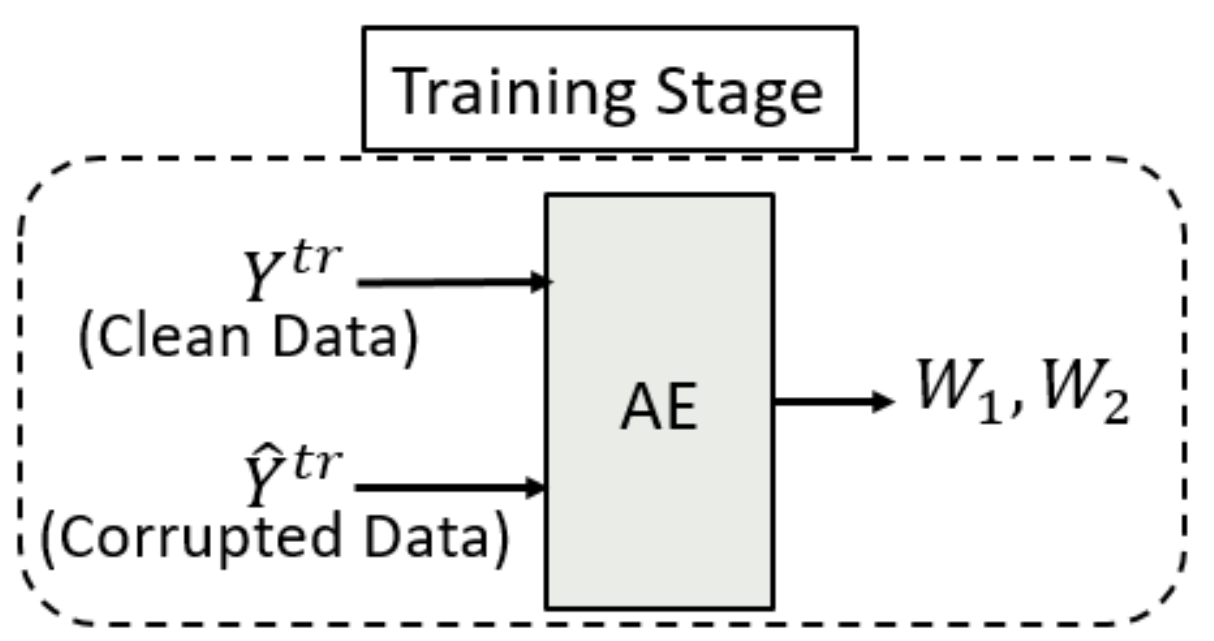}}\\
\hfil \subfigure[]{
\includegraphics[scale=0.5]{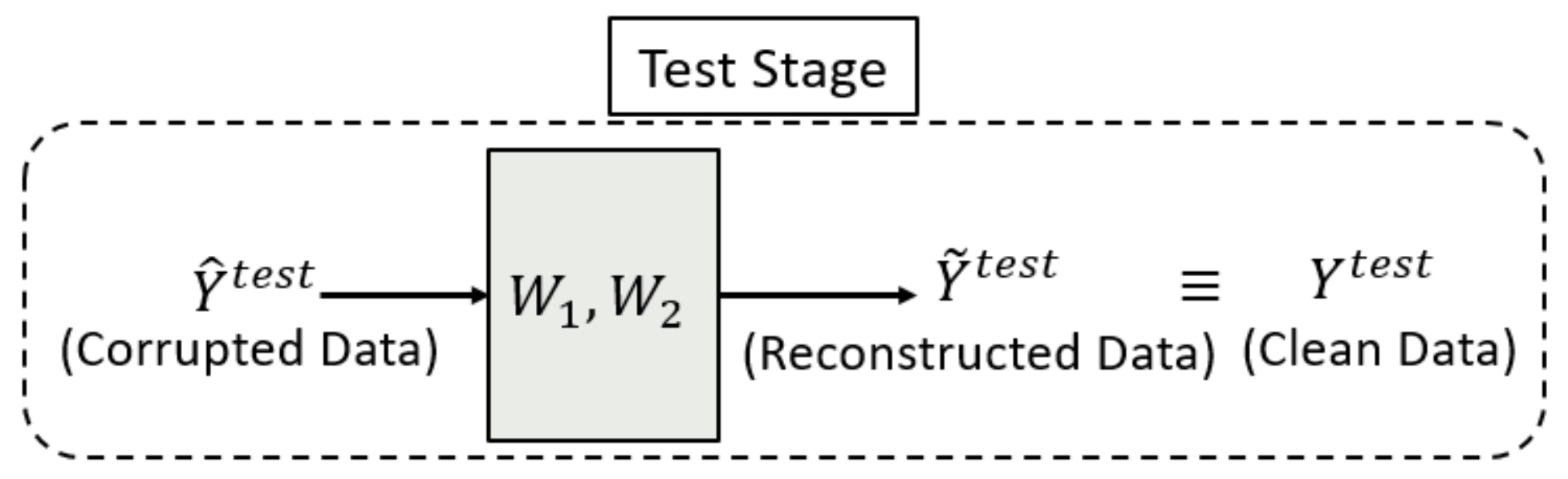}}
\hfil \caption{(a) Training phase, and (b) test phase, where measurements in through-wall
case $Y_{tr}$ as corrupted training radar images, $Y_{test}$ as corrupted test radar images and ones captured in free space $Y_{tr}^{clean}$, $Y_{test}^{clean}$ as clean training and test radar images.}
\label{fig:DE_training_test}
\end{figure}

We hypothesize that once the network is trained, we can use weight matrices $W_1$ and $W_2$ to obtain a denoised form $\tilde{Y}^{test}$ of the corrupted test data $\hat{Y}^{test}$ as shown in Fig.\ref{fig:DE_training_test}(b). 
\begin{equation}
\label{eqn:test}
    \tilde{Y}^{test}=W_2\phi(W_1\hat{Y}^{test})
\end{equation}
Note that the proposed denoising algorithm is significantly faster in generating denoised images at test time as it involves only a simple product operation in (\ref{eqn:test}). This makes the algorithm suitable for real-time applications.
\subsection{Metrics for evaluation}
We evaluate the effectiveness of the proposed clutter mitigation algorithm using two metrics- normalized mean square error (NMSE) and structural similarity index (SSIM). We consider the image captured in free space as the clean/ground truth image ($Y$). We calculate the NMSE and SSIM between the through-wall image $\hat{Y}^{test}$ and ground truth image before denoising (BD). Then we repeat the exercise after denoising (AD). In the second case, the NMSE and SSIM are calculated between the reconstructed / denoised image, $\tilde{Y}^{test}$, and the ground truth image. The hypothesis, here, is that the NMSE and SSIM will improve after denoising.

The NMSE is computed between $Y^{test}$ and $\tilde{Y}^{test}$ using (\ref{eqn:nmse}).
\begin{equation}
\label{eqn:nmse}
    NMSE=\frac{\left \| Y^{test}- \tilde{Y}^{test}\right \|_{2}^{2}}{\left \| Y^{test}\right \|_{2}^{2}}
\end{equation}
NMSE is sensitive to the energy of absolute errors of all the pixels of an image. However, NMSE between two images may be low even if they have drastically different structural features \cite{wang2002universal}. SSIM, on the other hand, is metric that provides information of the luminance ($L$), contrast ($C$) and structure difference ($S$), between the ground truth image $Y^{test}$,  and the test image $\tilde{Y}^{test}$. Its value should be $1$ if the images are identical. The overall measurement metric becomes the multiplicative combination of three measures shown in (\ref{eqn:SSIM})
\begin{equation}
\label{eqn:SSIM}
    SSIM(\hat{Y},Y)=[L(\hat{Y},Y)]^\alpha [C(\hat{Y},Y)]^\beta [S(\hat{Y},Y)]^\gamma 
\end{equation}
We assume $\alpha=\beta=\gamma=1$. The expressions for $L,C,S$ are 
\begin{equation}
\label{eqn:SSIM_L}
   L(\hat{Y},Y)=\frac{2\mu_{\hat{Y}}\mu_{Y}+C_1}{\mu_{\hat{Y}}^{2}+\mu_{Y}^{2}+C_1}
\end{equation}
\begin{equation}
\label{eqn:SSIM_C}
C(\hat{Y},Y)=\frac{2\sigma_{\hat{Y}}\sigma_{Y}+C_2}{\sigma_{\hat{Y}}^{2}+\sigma_{Y}^{2}+C_2}
\end{equation}
\begin{equation}
\label{eqn:SSIM_S}
S(\hat{Y},Y)=\frac{\sigma_{\hat{Y}Y}+C_3}{\sigma_{\hat{Y}}\sigma_{Y}+C_3}
\end{equation}
Here $\mu_{Y}$, $\mu_{\hat{Y}}$,$\sigma_{Y}$, $\sigma_{\hat{Y}}$ and $\sigma_{\hat{YY}}$ are the local means, standard deviations and the co-variance for the reference $Y$ and test images $\tilde{Y}$ respectively. Assuming $C_3=\frac{C_2}{2}$, the simplified index becomes.
\begin{equation}
\label{eqn:SSIM_Simple}
SSIM(\hat{Y},Y)=\frac{(2\mu_{\hat{Y}}\mu_{Y}+C_1)(2\sigma_{\hat{Y}Y}+C_2)}{(\mu_{\hat{Y}}^{2}+\mu_{Y}^{2}+C_1)(\sigma_{\hat{Y}}^{2}+\sigma_{Y}^{2}+C_2)}
\end{equation}
We therefore conclude that two images can be regarded as similar only when both NMSE is low and SSIM is close to 1.
\section{Simulation Results}
\label{Sec:SimExpSetUp}
In this section we describe the simulation method to generate a large database of Doppler enhanced frontal images of humans in diverse through-wall conditions. We adopt the technique described in \cite{ram2015high}. We only model the through-wall propagation phenomenology and do not consider multipath scattering from the ceiling, ground and lateral walls. The wall propagation phenomenology, modelled using finite difference time domain techniques (FDTD), and primitive based models of humans are hybridized to generate Doppler-enhanced frontal radar images. There may be considerable variations in the propagation conditions during training and test due to variations in the wall characteristics such as its dielectric constant and loss tangent. Modeling this diversity with independent FDTD simulations is computationally expensive. Therefore, we extend the simulation framework discussed in \cite{ram2010simulation} by incorporating stochasticity in the propagation channel using the stochastic FDTD (sFDTD) technique suggested by \cite{smith2012stochastic}. The sFDTD method introduces statistical variations in the electrical properties of the medium. The results of the simulations provide the mean and the variance estimates of the time-domain electromagnetic fields at every point in the problem space from which numerous samples of the through-wall propagation can be generated. We describe these steps in greater detail in the following section. 
\subsection{Stochastic Model of Through-Wall Propagation}
\begin{figure*}
\centering
\includegraphics[width=6.8in, height=2.0in]{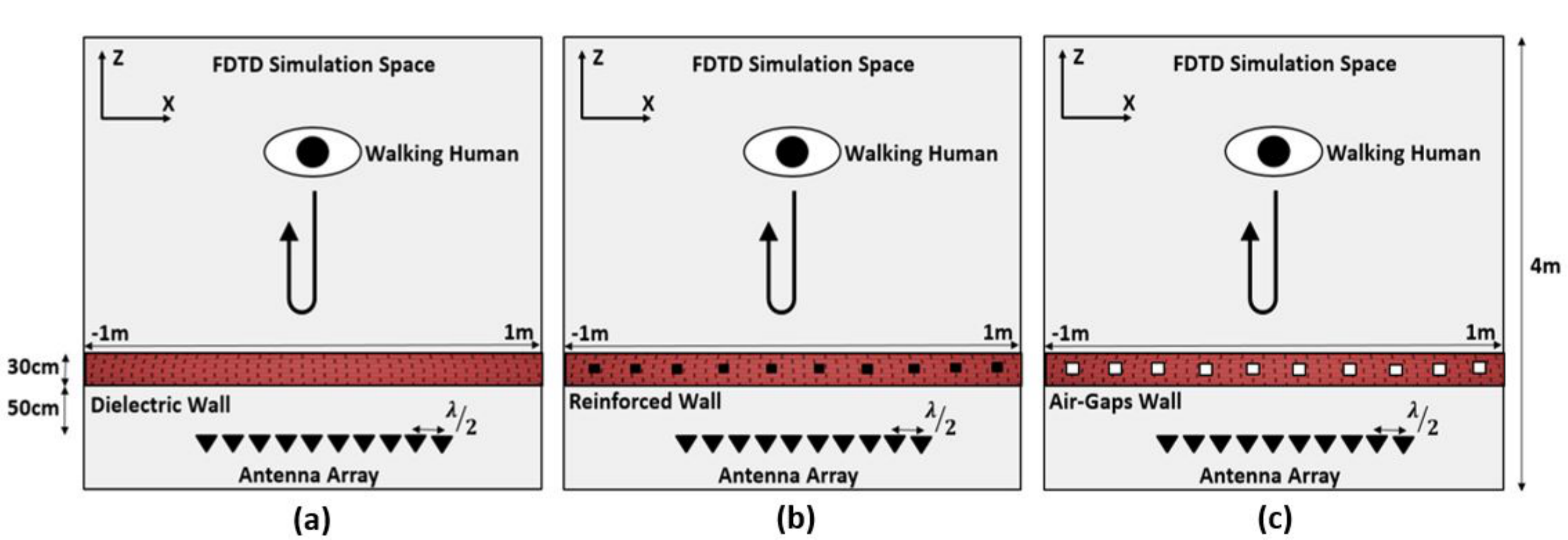}
\caption{Room Geometry in through-wall scenario (a) Dielectric (b) Reinforced wall and (c) Wall with air-gaps}
\label{fig:room_geometry}
\end{figure*}

As shown in Fig.\ref{fig:room_geometry}, we consider a two-dimensional simulation space extending from -1m to 1m and 0m to 4m along the $X$ and $Z$ directions respectively (assuming the problem is invariant along $Y$ height axis). The two-dimensional simulation framework is chosen to reduce the computational complexity of the problem and because most walls show homogeneity along the height. We assume that the radar consists of a 10 element uniform linear antenna array whose elements are spaced half wavelength apart. We independently simulate the excitation from each element of the array, located at $\vec{\rho}_s$, with a narrowband sinusoidal signal at 7.5GHz. 
The space is bounded by a perfectly matched layer and divided into spatial grids of size of $1/10^{th}$ the wavelength of the excitation source.  We considered three different wall configurations - a homogeneous dielectric wall (Fig.\ref{fig:room_geometry}(a)), a wall reinforced with metal rods (Fig.\ref{fig:room_geometry}(b)), and a wall with air gaps (Fig.\ref{fig:room_geometry}(c)). Each wall type is simulated independently. The dimensions of each wall is 2m x 30cm ($X$: -1m to 1m, $Z$: 1m to 1.3m). In all three cases, stochastic variations are introduced in the relative permittivity $\epsilon_{r}$ and conductivity $\sigma_{c}$ of each grid point in the wall. Therefore, even the single layer dielectric wall is not truly homogeneous. This is done to mirror real world conditions. 
For every point in space, $\vec{\rho}_p$, and at every time instant $t$, the sFDTD simulation gives the  mean time-domain electric field $\mu_E(\vec{\rho}_p, \vec{\rho}_s,t)$ and its standard deviation $\sigma_E(\vec{\rho}_p, \vec{\rho}_s,t)$. We use the Gaussian stochastic model to generate multiple samples of time-domain electric field values $E(\vec{\rho}_p, \vec{\rho}_s,t,\eta) \sim \mathcal{N}(\mu_E,\sigma_E^2)$ where each sample is denoted by $\eta$. The  $E(\vec{\rho}_p, \vec{\rho}_s,t,\eta)$ is fast Fourier transformed to obtain the corresponding frequency domain wall transfer function $H_{wall}(\vec{\rho}_p, \vec{\rho}_s,\eta)$ at 7.5GHz.

\begin{figure}[htbp]
\centering
\includegraphics[scale=0.42]{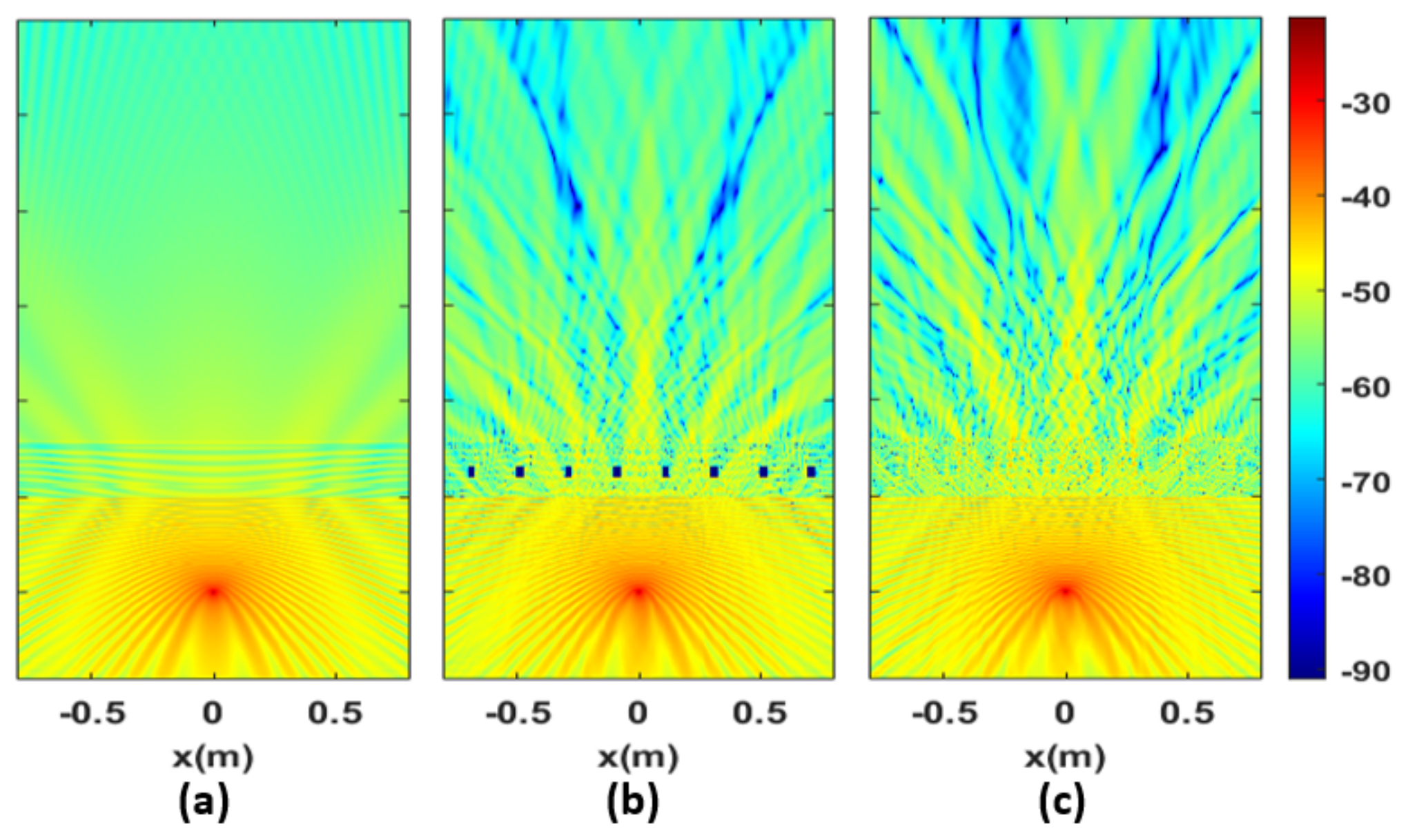}
\caption{Magnitude response at 7.5 GHz for (a)dielectric wall (b) reinforced dielectric wall, and (c) dielectric wall with air-gaps}
\label{fig:ques1}
\end{figure}

In a free space scenario, the magnitude of the electric field will decay as the distance from the source increases and the phase response will display well behaved circular wave-fronts emanating from a line source excitation. The propagation of a signal through a homogeneous dielectric wall undergoes a two-way attenuation of approximately 12dB when compared to free space. The magnitude response through a homogeneous dielectric wall. is shown in Figure~\ref{fig:ques1}(a). The phase response does not get perturbed much. 
However, in-homogeneous walls are complex propagation mediums. Therefore, the wall transfer functions, $H_{wall}(\vec{\rho}_p, \vec{\rho}_s,\eta)$, can introduce significant  phase  and  amplitude distortions  to  the  radar  signals. For example, Figure~\ref{fig:ques1}(b) and Figure~\ref{fig:ques1}(c) show the magnitude response of signal propagation through a reinforced wall and wall with air gaps when the source is located at $(0,0.5)m$. Both figures show that the inhomogeneity inside the wall causes multiple scattering that interfere destructively in some regions beyond the wall. The wall response is most severely distorted in the case of the wall with air-gaps.
\subsection{Modeling of electromagnetic radar scatter from dynamic humans in through-wall scenarios}
We  used  motion  capture  data  from  Sony  Computer  Entertainment  America  for realistically  modeling  human  animation. We considered a scenario where the human is facing the radar and walking towards it. The duration of the motion is 0.8s which corresponds to one complete human stride and data is sampled at 1KHz. Since our FDTD simulation spatial extent is limited, we removed translational motion of the human and only retained the dynamics of the swinging arms and legs. The human model is assumed to consist of multiple point scatterers located at $\vec{r}_b$ and of reflectivity $\sigma_b$ as given by \cite{ram2010simulation}. Therefore, the time-domain scattered returns at each $(m,n)$ antenna element is obtained by hybridizing $H_{wall}$ with the human scattering center model as shown in
\begin{eqnarray}
\begin{array} {lll}
\label{eqn:combined} 
  x(m,n,t,\eta)&=\Sigma_{b=1}^{N} \sqrt{\sigma_{b}(t)} \\
   &\times \left (Scale_{2D\rightarrow 3D} H_{wall}(\vec{\rho}_b(t), \vec{\rho}_s,\eta) \right )^2
\end{array}
\end{eqnarray}
Note that $H_{wall}(\vec{\rho}_p, \vec{\rho}_s,\eta)$, is the propagation factor from a two-dimensional source position $\vec{\rho}_s$ to field position $\vec{\rho}_b$ (projection of $\vec{r}_b$ in the two-dimensional space). 
We perform a suitable scaling operation to convert the two-dimensional propagation factor to three-dimensional transfer function between planar array $\vec{r}_{m,n}$ and field position $\vec{r}_b$ as discussed in \cite{ram2010simulation}.

The time domain radar data, $x(m,n,t)$, is simulated for a two-dimensional $10 \times 10$ planar array. The simulated radar parameters are listed in Table~\ref{table:sim_param}. The data is processed with three-dimensional Fourier transform to obtain $\chi(f_D,\theta,\phi)$. The Doppler processing is carried out using a coherent processing interval of 0.1s. Different body parts of a human move at different velocities and hence are resolved along the Doppler dimension. Doppler-enhanced frontal images are generated by integrating the responses of the peak scatterers at every Doppler. The incorporation of the additional Doppler dimension enables us to resolve multiple scatterers of the human along the two spatial dimensions \cite{ram2015doppler}.
\begin{table}[]
\centering
\caption{Simulated Radar Parameters}
\label{table:sim_param}
\begin{tabular}{|c|c|} \hline 
\textbf{Radar Parameters} & \textbf{Values} \\ \hline 
Radar Type & Narrowband \\ \hline 
Carrier frequency ($f_c$) & $7.5GHz$ \\ \hline 
Sampling frequency ($f_s$) & $1000Hz$ \\ \hline 
Integration time ($T$) & $0.8s$ \\ \hline 
Dwell time or short time ($t_D$) & $0.1s$ \\ \hline 
Maximum Doppler ($f_{Dmax}$) & $\pm 500Hz$ \\ \hline 
Doppler resolution (${\Delta f}_D$) & $10Hz$ \\ \hline 
Number of antenna elements ($N\times N)$ & $10\times 10$ \\ \hline 
Azimuth Beamwidth (${\mathrm{\Delta }\phi }_{azi}$) & ${10}^{{}^\circ }$ \\ \hline 
Azimuth Beamwidth (${\mathrm{\Delta }\theta }_{ele}$) & ${10}^{{}^\circ }$ \\ \hline 
Field of View (${\phi }_{azi})$ & $-{90}^{{}^\circ }\ to\ {90}^{{}^\circ }$ \\ \hline 
Field of View (${\theta }_{azi})$ & $-{90}^{{}^\circ }\ to\ {90}^{{}^\circ }$ \\ \hline 
\end{tabular}
\end{table}
\subsection{Simulation Results and Analyses}
\label{Sec:SimResults}
We simulate radar returns from a walking human at $7.5$GHz, in free space, and then repeat the exercise for different through-wall scenarios as discussed above. As a result, we obtain clean free space radar images of the moving human as well as corrupted through-wall radar images. 
\begin{figure*}[htbp]
\centering 
\includegraphics[scale=0.62]{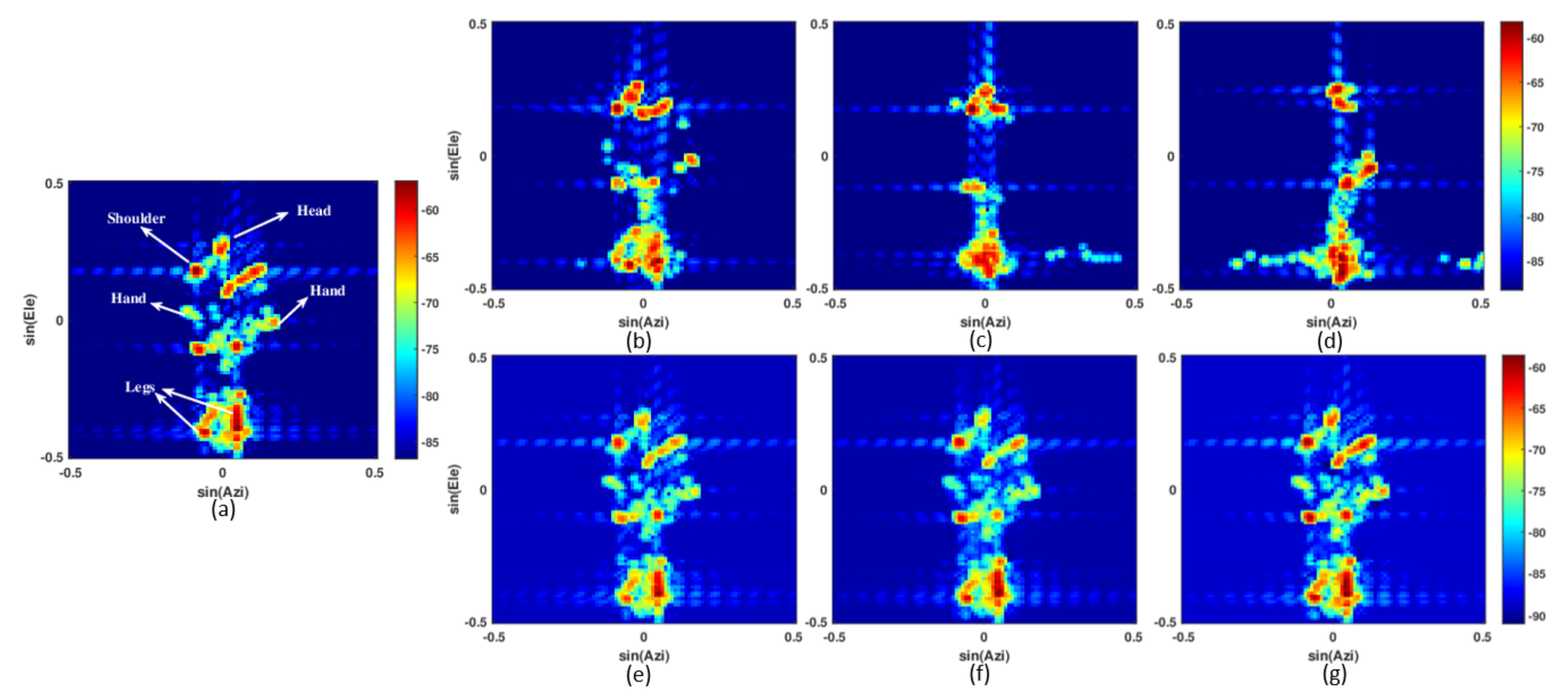}
\hfil \\
\caption{Simulated Doppler enhanced frontal image of a walking human in (a) Free space, (b),(c),(d) through wall scenarios-dielectric Wall, reinforced dielectric wall, dielectric wall with air gaps respectively. (e),(f),(g) Denoised images  of a walking human in through wall scenarios-dielectric Wall, reinforced dielectric wall, dielectric wall with air gaps respectively using the proposed algorithm.}
\label{fig:simulation_model_throughwall}
\end{figure*}
Fig.\ref{fig:simulation_model_throughwall}(a) shows the Doppler enhanced frontal image in the free space condition. The image falls within the $\pm 30^{\circ}$ field-of-view of the radar along elevation and azimuth. We can clearly discern both arms, legs and head of the human in the image. There is some smearing near the legs due to the limited resolution along azimuth and elevation of the array. Imaging on the other hand is adversely affected when captured in through-wall conditions due to the multipath.
Fig.\ref{fig:simulation_model_throughwall}(b)-(d) show the frontal images in the dielectric wall, reinforced wall and wall with air-gaps scenarios respectively. Firstly two-way propagation through the wall suffers an attenuation of approximately 12 dB. Hence, the strength of some of the peak scatterers become too weak to be visible on the same dynamic scale as that of free space case. Additionally positions of few scatterers get displaced along the azimuth direction due to refraction. The images of humans behind the metal reinforced wall and wall with air gaps are considerably more distorted. Some of the point scatterers are not visible at all because these lie at regions of destructive interference as shown in Figure.\ref{fig:ques1}b and c. We apply the denoising algorithm discussed in Section \ref{Sec:Theory} to the simulation data and evaluate its performance at denoising the corrupted images. Reconstructed images after the denoising step are shown in Fig. \ref{fig:simulation_model_throughwall}(e)-(f). Here, the reconstructed images look quite similar to the ground truth - that is the free space image shown in Fig. \ref{fig:simulation_model_throughwall}(a).
\begin{table*}[]
\centering
\caption{Denoising results between clean and corrupted Doppler enhanced frontal images
for different through-wall conditions.SSIM:between
corrupted and free space image before denoising (BD) and SSIM :between reconstructed and free space image after denoising (AD)}
\label{table:Simulation_results_num_frames_ssim}
\begin{tabular}{|l|c|c|l|l|l|l|l|}
\hline
\multicolumn{2}{|c|}{\multirow{2}{*}{\textbf{Wall Scenario}}}                                             & \multirow{2}{*}{\textbf{\begin{tabular}[c]{@{}c@{}}Denoising Metric\\ (SSIM)\end{tabular}}} & \multicolumn{5}{c|}{\textbf{Number of Frames}}                                                                                                                             \\ \cline{4-8} 
\multicolumn{2}{|c|}{}                                                                                    &                                                                                             & \multicolumn{1}{c|}{\textbf{1}} & \multicolumn{1}{c|}{\textbf{5}} & \multicolumn{1}{c|}{\textbf{10}} & \multicolumn{1}{c|}{\textbf{20}} & \multicolumn{1}{c|}{\textbf{30}} \\ \hline\hline
\multirow{6}{*}{\textbf{Same Wall}} & \multirow{2}{*}{\textbf{Dielectric}}                              & \textbf{BD}                                                                                 & 0.09                            & 0.55                            & 0.63                             & 0.67                             & 0.64                             \\ \cline{3-8} 
                                      &                                                                   & \textbf{AD}                                                                                 & 0.43                            & 0.87                            & 0.90                             & 0.89                             & 0.86                             \\ \cline{2-8} \cline{2-8}
                                      & \multirow{2}{*}{\textbf{Reinforced}}                              & \textbf{BD}                                                                                 & 0.05                            & 0.24                            & 0.46                             & 0.55                             & 0.47                             \\ \cline{3-8} 
                                      &                                                                   & \textbf{AD}                                                                                 & 0.43                            & 0.81                            & 0.87                             & 0.84                             & 0.80                             \\ \cline{2-8} \cline{2-8}
                                      & \multicolumn{1}{l|}{\multirow{2}{*}{\textbf{Wall With Air-gaps}}} & \textbf{BD}                                                                                 & 0.03                            & 0.01                            & 0.34                             & 0.51                             & 0.43                             \\ \cline{3-8} 
                                      & \multicolumn{1}{l|}{}                                             & \textbf{AD}                                                                                 & 0.42                            & 0.82                            & 0.83                             & 0.82                             & 0.77                             \\ \hline\hline
\multicolumn{2}{|c|}{\multirow{2}{*}{\textbf{Different Walls}}}                                            & \textbf{BD}                                                                                 & 0.04                            & 0.26                            & 0.43                             & 0.52                             & 0.45                             \\ \cline{3-8} 
\multicolumn{2}{|c|}{}                                                                                    & \textbf{AD}                                                                                 & 0.19                            & 0.81                            & 0.84                             & 0.83                             & 0.78                             \\ \hline
\end{tabular}
\end{table*}

\begin{table*}[]
\centering
\caption{Denoising results between clean and corrupted Doppler enhanced frontal images
for different through-wall conditions.NMSE:between
corrupted and free space image before denoising (BD) and NMSE :between reconstructed and free space image after denoising (AD)}
\label{table:Simulation_results_num_frames_nmse}
\begin{tabular}{|l|c|c|l|l|l|l|l|}
\hline
\multicolumn{2}{|c|}{\multirow{2}{*}{\textbf{Wall Scenario}}}                                             & \multirow{2}{*}{\textbf{\begin{tabular}[c]{@{}c@{}}Denoising Metric\\ (NMSE)\end{tabular}}} & \multicolumn{5}{c|}{\textbf{Number of Frames}}                                                                                                                             \\ \cline{4-8} 
\multicolumn{2}{|c|}{}                                                                                    &                                                                                             & \multicolumn{1}{c|}{\textbf{1}} & \multicolumn{1}{c|}{\textbf{5}} & \multicolumn{1}{c|}{\textbf{10}} & \multicolumn{1}{c|}{\textbf{20}} & \multicolumn{1}{c|}{\textbf{30}} \\ \hline
\multirow{6}{*}{\textbf{Same Wall}} & \multirow{2}{*}{\textbf{Dielectric}}                              & \textbf{BD}                                                                                 & 3.60                            & 3.50                            & 3.18                             & 3.49                             & 3.39                             \\ \cline{3-8} 
                                      &                                                                   & \textbf{AD}                                                                                 & 1.05                            & 0.85                            & 0.99                             & 0.92                             & 1.56                             \\ \cline{2-8} \cline{2-8}
                                      & \multirow{2}{*}{\textbf{Reinforced}}                              & \textbf{BD}                                                                                 & 5.71                            & 5.91                            & 5.57                             & 5.60                             & 5.01                             \\ \cline{3-8} 
                                      &                                                                   & \textbf{AD}                                                                                 & 1.60                            & 1.09                            & 1.08                             & 1.16                             & 1.36                             \\ \cline{2-8}\cline{2-8} 
                                      & \multicolumn{1}{l|}{\multirow{2}{*}{\textbf{Wall With Air-gaps}}} & \textbf{BD}                                                                                 & 4.42                            & 4.01                            & 3.50                             & 3.25                             & 2.86                             \\ \cline{3-8} 
                                      & \multicolumn{1}{l|}{}                                             & \textbf{AD}                                                                                 & 1.14                            & 1.18                            & 1.36                             & 1.32                             & 1.32                             \\ \hline
\multicolumn{2}{|c|}{\multirow{2}{*}{\textbf{Different Walls}}}                                            & \textbf{BD}                                                                                 & 5.01                            & 4.49                            & 4.07                             & 4.13                             & 3.78                             \\ \cline{3-8} 
\multicolumn{2}{|c|}{}                                                                                    & \textbf{AD}                                                                                 & 0.71                            & 1.16                            & 1.26                             & 1.60                             & 1.58                             \\ \hline
\end{tabular}
\end{table*}

The results are obtained by optimizing the number of nodes ($r$) in the hidden layer and the mapping function connecting the input and the hidden layer. We fixed the hidden layer dimension of the autoencoder network to be 500 and the mapping to be linear between input and the hidden layer. The choice of these parameters are discussed in the appendix \ref{appendixxa}.  $80\%$ of the simulated data are used for training and the remaining for test. During training, the weight matrices $W_1$ and $W_2$ each of size $[500 \times 8464]$ and $[8464 \times 500]$ respectively are first randomly initialized. Here, 8464 denotes the number of pixels in the image. The weights are updated over successive iterations as discussed in the previous section. The regularization parameter is chosen to be 1. Once learned, the weights are used for test. 

\begin{figure}[t]
\centering
\includegraphics[scale=0.35]{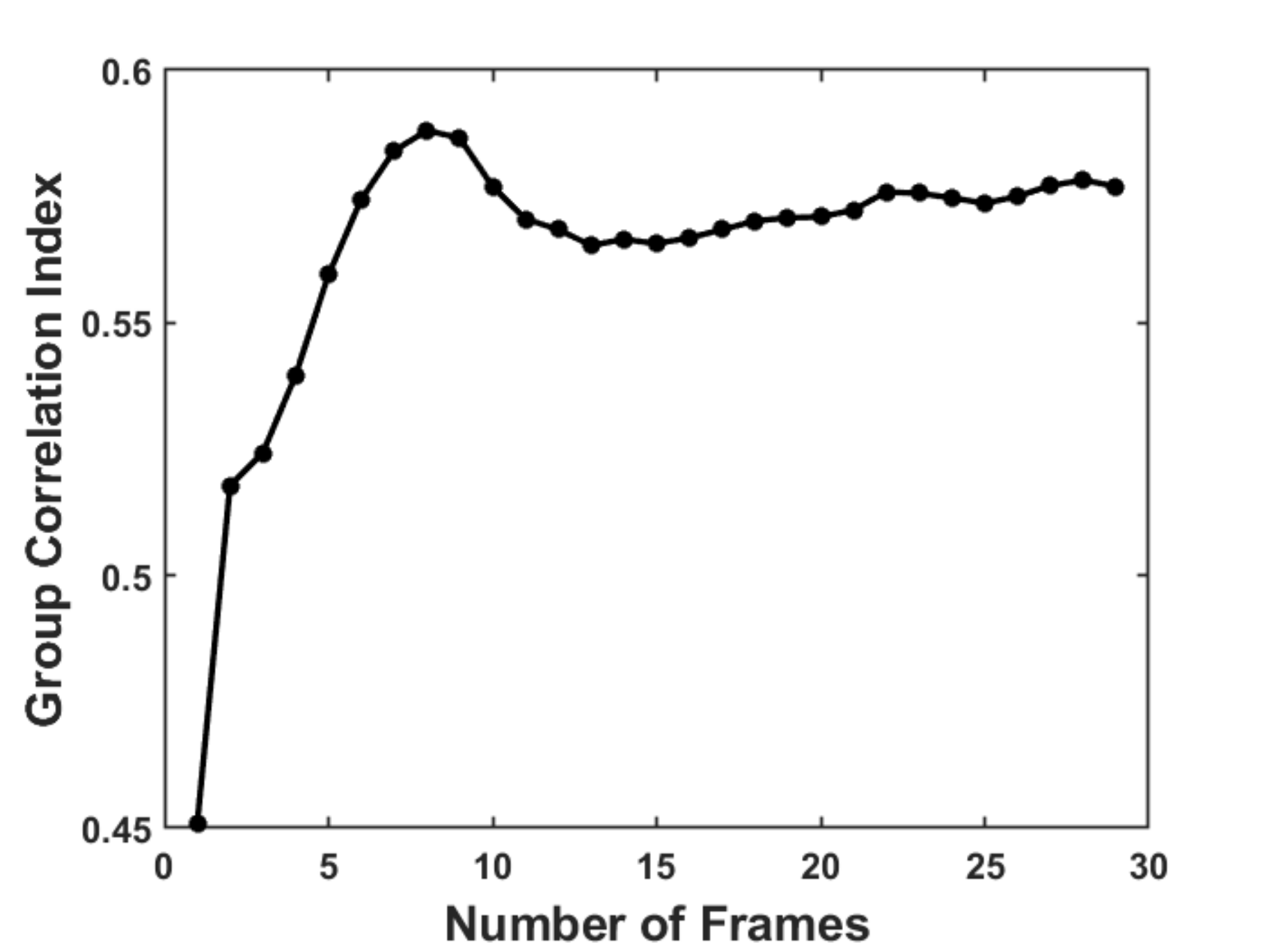}
\caption{Group correlation across multiple frames}
\label{fig:ques2}
\end{figure}

Table. \ref{table:Simulation_results_num_frames_ssim}$-$\ref{table:Simulation_results_num_frames_nmse} show the results for two scenarios using the metrics, SSIM and NMSE, as a function of number of distinct frames of the human walking motion respectively. We compare the metrics obtained from images generated \emph{before denoising (BD)} with those obtained \emph{after denoising (AD)}.  First, we consider the scenario, where the autoencoder is trained with data from a specific wall configuration and then subsequently tested on images generated from the same wall configuration. Note that even in the \emph{same wall} case, there is diversity in the training and test data due to the statistical variations in the wall parameters. Before denoising, the dielectric wall case has lowest error when compared to the reinforced and air-gaps walls. This is because the quality of the radar images are a function of the phase and amplitude distortions introduced by the walls to the radar signals. Therefore, based on the magnitude and phase responses shown in Fig.\ref{fig:ques1}(a) and (b), we observe the results deteriorate most in the case of the wall with air-gaps.  The error between reconstructed and the free space images drop significantly for all wall types after passing through the denoising network. 
We varied the number of frames from $1$ to $30$ to increase the diversity in the human motions. Now, since this is a continuous motion, there is some degree of correlation between images obtained from the consecutive frames. This is reflected in the group correlation index across multiple frames shown in Fig.~\ref{fig:ques2}. The group correlation increases until 10 frames after which there is no further improvement. Hence, the performance seems to improve when we increase from a single frame to 10 number of frames in the Tables. \ref{table:Simulation_results_num_frames_ssim}$-$\ref{table:Simulation_results_num_frames_nmse} as the training data captures the diversity of motions. Beyond this, the performance of the denoising algorithm slightly deteriorates due to the diversity of the images due to the continuous motion.  Also note that NMSE and SSIM do not behave in an identical manner for all the cases as they indicate different aspects of similarity of images. 

Next, we study the scenario when the network is trained and tested on images captured from the different through-wall scenarios. This is a significantly more challenging scenario since no information of the type of wall is available during the test phase. The results, however, show a very good performance (NMSE and SSIM) comparable to that of the same wall scenarios. 
\subsection{Impact of radar-target aspect angle}
\label{Sec:SimResults_multi_aspect}
In order to understand the generality of the proposed denoising solution, we trained our autoencoder network with human radar images captured at different aspect angles with respect to radar line-of-sight conditions. We analyzed the performance of our algorithm for four aspect angles- $0^{\circ}$, $45^{\circ}$, $90^{\circ}$, $180^{\circ}$. Here $0^{\circ}$ aspect angle means the person is walking towards the radar and $180^{\circ}$ means the person is walking away from the radar. Similarly $90^{\circ}$ corresponds to the motion along the tangential direction to the radar. We studied the efficacy of the algorithm for reinforced wall which, as mentioned earlier, is one of the most complex walls. We tested the performance of the denoising autoencoder on two scenarios: In the first scenario, both the training and test data are gathered at the same aspect angle (identical training and test scenario); In the second scenario,  we used images captured at multiple different aspect angles for both training and testing the autoencoder. Therefore, during test, the algorithm is not provided information of the aspect angle of the data. We used both NMSE and SSIM to measure the performance and report the results in  Table~\ref{table:sim_results_num_frames_aspect_angles}.
\begin{table}[]
\centering
\caption{Denoising results between clean and corrupted images (captured behind reinforced wall) for different aspect angles. SSIM,NMSE:between
corrupted and free space image before denoising (BD) and SSIM,NMSE:
between reconstructed and free space image after denoising (AD)}
\label{table:sim_results_num_frames_aspect_angles}
\begin{tabular}{|c|c|c|c|c|}
\hline
\multicolumn{3}{|c|}{\multirow{2}{*}{\textbf{\begin{tabular}[c]{@{}c@{}}Wall Scenario\\ (Reinforced)\end{tabular}}}}                 & \multicolumn{2}{c|}{\textbf{Denoising Metric}} \\ \cline{4-5} 
\multicolumn{3}{|c|}{}                                                                                                                  & \textbf{SSIM}          & \textbf{NMSE}         \\ \hline
\multirow{10}{*}{\textbf{\begin{tabular}[c]{@{}c@{}}Aspect Angle\\ \\ (Degree)\end{tabular}}} & \multirow{2}{*}{\textbf{0}}                                & \textbf{BD}                      & 0.05                   & 5.64                  \\ \cline{3-5} 
                                        &                                                            & \textbf{AD}                      & 0.64                   & 1.16                  \\ \cline{2-5} 
                                        & \multirow{2}{*}{\textbf{45}}                               & \textbf{BD}                      & 0.36                   & 4.01                  \\ \cline{3-5} 
                                        &                                                            & \textbf{AD}                      & 0.73                   & 1.02                  \\ \cline{2-5} 
                                        & \multirow{2}{*}{\textbf{90}}                               & \textbf{BD}                      & 0.01                   & 4.33                  \\ \cline{3-5} 
                                        &                                                            & \textbf{AD}                      & 0.60                   & 1.67                  \\ \cline{2-5} 
                                        & \multirow{2}{*}{\textbf{180}}                              & \multicolumn{1}{l|}{\textbf{BD}} & 0.46                   & 5.01                  \\ \cline{3-5} 
                                        &                                                            & \multicolumn{1}{l|}{\textbf{AD}} & 0.80                   & 1.23                  \\ \cline{2-5} 
                                        & \multicolumn{1}{l|}{\multirow{2}{*}{\textbf{0,45,90,180}}} & \multicolumn{1}{l|}{\textbf{BD}} & 0.41                   & 4.60                  \\ \cline{3-5} 
                                        & \multicolumn{1}{l|}{}                                      & \multicolumn{1}{l|}{\textbf{AD}} & 0.71                   & 1.44                  \\ \hline
\end{tabular}
\end{table}
We observe highest error when the algorithm is trained with data captured at $90^{\circ}$ aspect angle that is when the human walks in a direction tangential to the radar. This is most likely because of the inherent distortions in these frontal images due to the limited separation of point scatterers on the subject along the azimuth direction. Likewise, the Dopplers of the different point scatterers on the human body are not well resolved due to the tangential motion. The results reported for all the aspect angles show significant improvement after denoising. When we consider data from multiple aspect angles, the denoising significantly helps in reconstructing images close to free space images even when the algorithm is not provided any information of the exact aspect angle at which a person is walking. Therefore we can infer that the autoencoder is specifically suited for problems dealing with a great deal of diversity in the target and channel conditions. It can significantly denoise (i) images captured in similar and dissimilar- wall conditions as well as (ii) images captured at different aspect angles of the target provided there is sufficient diversity across training data.
\section{Measurement Results}
\subsection{Measurement Data Collection}
\label{Sec:Measurement_setup}
In this section, we evaluate the performance of our algorithm using wideband measurement data captured in both free space and through-wall conditions. The data is collected using Walabot Pro \cite{walabot_2017}, a wideband(3.3-10.3GHz) 3D-programmable RF imaging sensor. Walabot is a low power uncalibrated sensor with limited range in through-wall scenarios. It uses a $4 \times 4$ antenna array to illuminate the area in front of it to capture the back-scattered signals. The hardware radar parameters are listed in Table.\ref{table:meas_param}. The wideband radar data are processed with three-dimensional Fourier transform to obtain range-azimuth-elevation results. The peak scatterers across all the range gates are coherently summed to obtain range enhanced frontal images of the targets. The assumption here, is that the targets are still or slow moving.
\begin{figure*}[htbp]
\centering 
\includegraphics[scale=0.55]{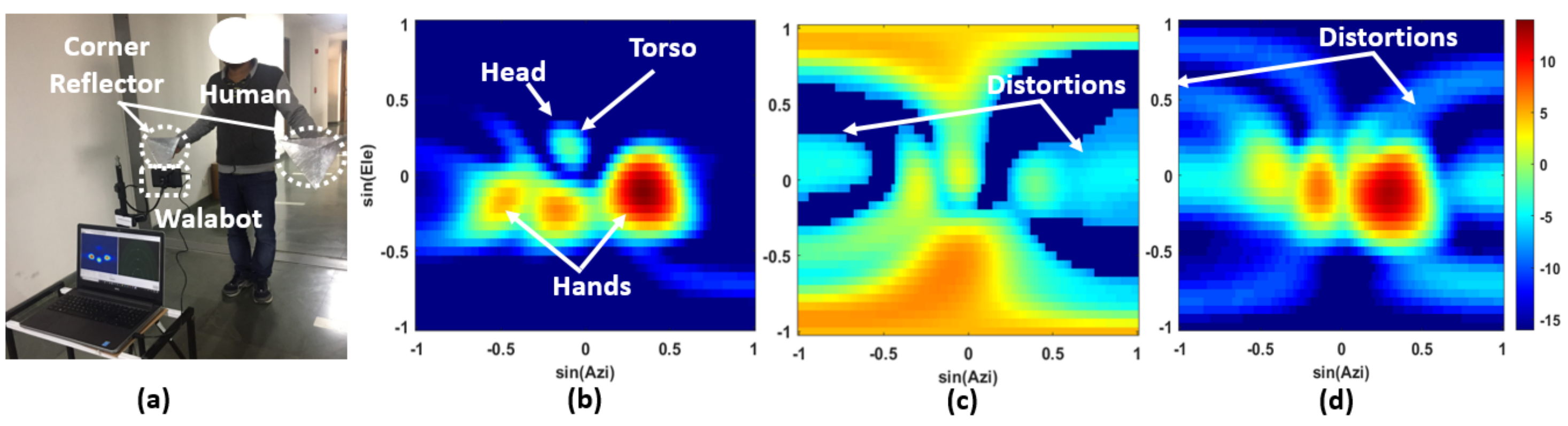}
\hfil \caption{(a) Measurement setup in free space and measured range enhanced frontal image of a human subject in (b) free space, (c),(d) behind a glass wall, wood wall respectively }
\label{fig:ques4}
\end{figure*}

 Our clean measurement data consist of range enhanced frontal images of a human gathered in line-of-sight conditions in an environment mostly free of clutter. The through-wall measurement data comprise of images captured through a 2cm thick glass wall and a 3cm thick wooden wall. The subject stands in front of the radar at a standoff distance of 2m carrying two corner reflectors covered with aluminum tape to enhance the reflectivity from the hands as shown in Fig~\ref{fig:ques4}(a). Therefore, the target is an extended target with multiple point scatterers. The experiments are performed on 4 human subjects of different heights and girth at different orientations ($-45^{\circ} to +45^{\circ}$) with respect to the radar but always facing the radar. For each of these subjects, we captured 75 measurements. An example of the resultant radar image in free space is shown in Figure~\ref{fig:ques4}(b) where we can clearly discern the torso, legs and two arms of the human. Radar images are corrupted when the measurements are gathered under different through-wall conditions. Some examples of the distortions are presented in Fig.~\ref{fig:ques4}. Fig.~\ref{fig:ques4}(c)-(d) correspond to images in through-glass wall and through-wood wall conditions respectively. These images are considerably distorted due to the complex interaction between the wall and the target. 
\begin{table}[]
\centering
\caption{Parameters of Real Radar Setup, *- values derived from available information}
\label{table:meas_param}
\begin{tabular}{| c | c |} \hline 
\textbf{Radar Parameters} & \textbf{Values} \\ \hline 
Radar Type & Broadband \\ \hline 
Bandwidth & $3.3GHz-10.3GHz$ \\ \hline 
Maximum Range ($R_{max}$) & $10m$* \\ \hline 
Range resolution ($\Delta r$) & $0.2m$* \\ \hline 
ADC & $8\ bit$ \\ \hline 
Number of antenna elements ($N\times N)$ & $4\times 4$ \\ \hline 
Azimuth Beamwidth (${\mathrm{\Delta }\phi }_{azi}$) & ${25}^{{}^\circ }$* \\ \hline 
Azimuth Beamwidth (${\mathrm{\Delta }\theta }_{ele}$) & ${25}^{{}^\circ }$* \\ \hline 
Field of View (${\phi }_{azi})$ & $-{90}^{{}^\circ }\ to\ {90}^{{}^\circ }$* \\ \hline 
Field of View (${\theta }_{azi})$ & $-{90}^{{}^\circ }\ to\ {90}^{{}^\circ }$* \\ \hline 
\end{tabular}
\end{table}
The size of each radar image is $[91 \times 37]$ pixels. The image is vectorised to obtain a $[3367 \times 1]$ vector. 80\% of the measurement data are used for training, along with corresponding free-space images, and the remaining used for testing. Once trained, the weight matrices $W_1$ and $W_2$ are used to denoise the corrupted test images using the equation (\ref{eqn:test}). Analogous to simulations, we examine the variation of denoising performance for different number of nodes in the hidden layer, for different mapping functions (linear, non-linear- tanh and sigmoid) and for different proportions of training to test data.

\subsection{Measurement Results and Analyses}
\label{Sec:MeasurementResults}
  We tested the performance of the denoising autoencoder for \emph{same wall} and \emph{different wall} scenarios. In the same wall scenario, both the training and test data are gathered from the same type of wall. In the different wall scenario, data from multiple walls are used for training the autoencoder which is subsequently used for denoising images from any of the two walls. The reconstruction results are presented as a function of percentage of training data to test data in Table~\ref{table:Measurement_human_results}.  These results have been obtained using an autoencoder where the hidden layer has 1500 nodes and the mapping function is sigmoid. The choice of these parameters are discussed in the appendix. The table shows SSIM and NMSE between the denoised radar images in through-wall and corresponding radar image gathered in free space conditions. We compare the metrics before denoising (BD) with those after denoising (AD). We observe there is significant improvement in SSIM and reduction of NMSE after denoising. The performance improves as the percentage of training to test data increases for both the same wall and for different wall scenarios. In other words, the performance during test relies on adequate training data. The error for the different wall scenario is only slightly higher than the same wall scenario. This is the scenario when the test algorithm has no knowledge of the wall scenario. Note that in the case of the wideband measurements, we have not presented the result as a function of the number of frames. This is because, the targets are static and each measurement is independent with no correlation between them. 
\begin{table*}[]
\centering
\caption{Denoising results between clean and corrupted measurement images of real humans
for different through-wall conditions under varying percentage of training data. NMSE, SSIM :between
corrupted and free space image before denoising (BD) and NMSE,SSIM:
between reconstructed and free space image after denoising (AD)}
\label{table:Measurement_human_results}
\begin{tabular}{|c|l|l|c|c|c|c|c|c|c|c|}
\hline
\multicolumn{3}{|c|}{\multirow{3}{*}{\textbf{Wall Scenario}}}                                                                         & \multicolumn{8}{c|}{\textbf{Denoising Metric}}                                                                                                                                                          \\ \cline{4-11} 
\multicolumn{3}{|c|}{}                                                                                                                & \multicolumn{4}{c|}{\textbf{\begin{tabular}[c]{@{}c@{}}SSIM\\ (\% of Training Data)\end{tabular}}} & \multicolumn{4}{c|}{\textbf{\begin{tabular}[c]{@{}c@{}}NMSE\\ (\% of Training Data)\end{tabular}}} \\ \cline{4-11} 
\multicolumn{3}{|c|}{}                                                                                                                & \textbf{20}             & \textbf{40}            & \textbf{60}            & \textbf{80}            & \textbf{20}             & \textbf{40}            & \textbf{60}            & \textbf{80}            \\ \hline
\multicolumn{1}{|l|}{\multirow{4}{*}{\textbf{Same Wall}}} & \multirow{2}{*}{\textbf{Glass Wall}} & \textbf{BD}                      & 0.20                    & 0.20                   & 0.22                   & 0.21                   & 36.09                   & 38.47                  & 33.52                  & 38.70                  \\ \cline{3-11} 
\multicolumn{1}{|l|}{}                                      &                                      & \textbf{AD}                      & 0.53                    & 0.70                   & 0.88                   & 0.97                   & 7.81                    & 5.10                   & 4.68                   & 3.69                   \\ \cline{2-11} 
\multicolumn{1}{|l|}{}                                      & \multirow{2}{*}{\textbf{Wood Wall}}  & \textbf{BD}                      & 0.22                    & 0.22                   & 0.23                   & 0.23                   & 28.21                   & 29.04                  & 27.51                  & 33.76                  \\ \cline{3-11} 
\multicolumn{1}{|l|}{}                                      &                                      & \textbf{AD}                      & 0.50                    & 0.55                   & 0.82                   & 0.91                   & 7.59                    & 7.10                   & 4.59                   & 4.24                   \\ \hline
\multicolumn{2}{|c|}{\multirow{2}{*}{\textbf{Different Walls}}}                                     & \multicolumn{1}{c|}{\textbf{BD}} & 0.23                    & 0.22                   & 0.23                   & 0.22                   & 29.00                   & 28.63                  & 28.43                  & 29.49                  \\ \cline{3-11} 
\multicolumn{2}{|c|}{}                                                                             & \multicolumn{1}{c|}{\textbf{AD}} & 0.46                    & 0.59                   & 0.70                   & 0.89                   & 8.40                    & 7.39                   & 6.19                   & 4.97                   \\ \hline
\end{tabular}
\end{table*}
\section{Discussion on Results}
\subsection{Computational Complexity Evaluation}
\begin{figure}[htbp]
\centering 
\includegraphics[scale=0.45]{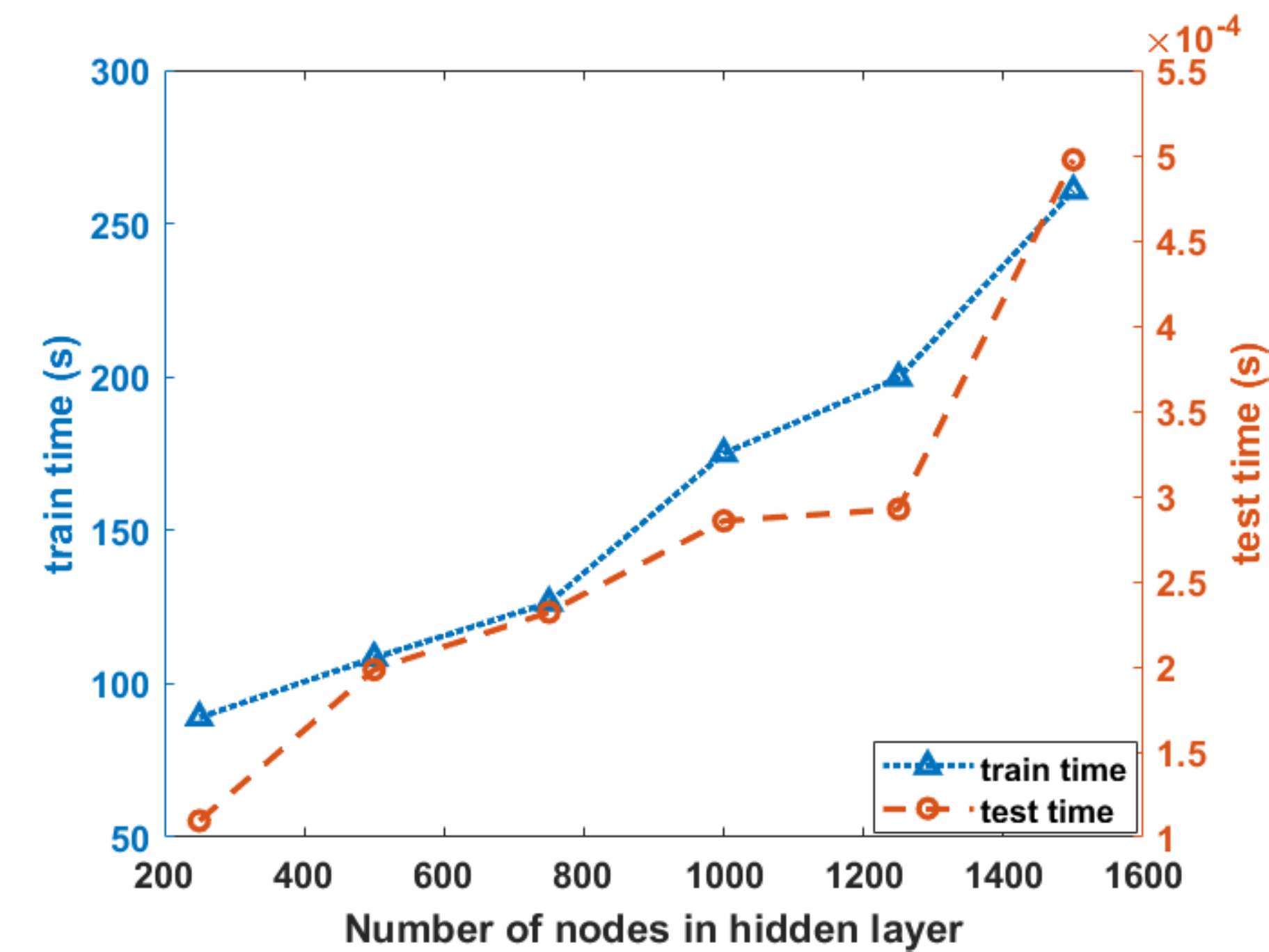}
\hfil \caption{Computational time as a function of number of nodes in the hidden layer for (left y-axis) training phase (right y-axis) test phase}
\label{fig:ques6_2nd}
\end{figure}

The real-time performance of the algorithm relies on the test time and the test memory rather than training time. During test, we perform matrix multiplication operations of the trained weights $W_1$ and $W_2$ with test image $X_{test}$. The sizes of the weight matrices and the image matrix are $r \times N$, $N \times r$ and $N \times 1$ respectively, where $N$ denotes the number of pixels in the image and $r$ denotes the number of hidden nodes in the autoencoder such that the number of nodes is always well below the number of pixels. The computational complexity therefore is $\bigO(rN)$. We ran our algorithm on  Matlab 2015b, where all the variables were stored as 64 bit floats, with an Intel(R) Core(TM) i7-5500 processor running at 2.40 GHz. We report the test and training times of our algorithms as a function of the number of nodes of the hidden layers in Figure~\ref{fig:ques6_2nd}.  Both the training and test times are higher when the hidden layer dimensionality is more. The test time is significantly low even for the highest number of hidden nodes (1500). The computational memory in all of these cases was less than 500MB.  
Therefore, these test operations can be carried out in easily available processors such as Raspberry PI 3+ (with a 1GB RAM and 1.4GHz clock speed). 
\subsection{Diversity of training data}
The training data must be sufficiently large to handle the diversity of target conditions, channel conditions and any type of labelling errors between free space and through-wall images.\\
\textbf{Diversity of target data:} In our work, our autoencoder has been trained to handle the diversity in the size, shape and orientation or aspect angle of the target with respect to the radar. In the case of dynamic motions, the correlation between consecutive frames facilitated in improving the denoising performance.    

\noindent\textbf{Diversity of channel data:} Next, the proposed approach does not require the knowledge of the exact wall conditions or analytical framework during the test phase. Instead, the algorithm was capable of denoising images obtained from diverse through-wall conditions. 

\noindent\textbf{Labelling errors between free-space and through-wall images:} Finally, in practice, it may be nearly impossible to gather correlated images in free space and through-wall conditions especially for dynamic targets. For example, it may not be possible to replicate human motions in two different scenarios. Therefore, the algorithm must tolerate some degree of diversity in the motion characteristics during test and training phases. A sensitivity analysis of mismatch/labeling error between clean (free space) and the corrupted (through-wall) training images is not considered in this work. Generally in machine learning scenarios, these algorithms are quite robust to reasonable random errors in the training set arising due to incorrectly labelled data. However, the algorithms are less robust to systematic errors when the samples are consistently incorrectly labelled.    
\section{Conclusion}
\label{Sec:Conclusion}
We demonstrate the efficacy of the denoising autoencoder network at mitigating the distortions and clutter introduced by wall propagation on radar images. The proposed approach requires neither prior information of the wall characteristics nor any kind of analytic framework to describe the wall propagation effects. Instead, the algorithm relies on the availability of a huge training data set comprising of distorted radar images captured in diverse through-wall scenarios and the corresponding clean images in line-of-sight conditions. Once trained, the algorithm is capable of mitigating through-wall effects of similar walls though not necessarily identical walls. This capability makes this approach suitable for tracking humans under diverse propagation environments. We evaluated the performance of the algorithm on both static and dynamic targets. The radar images of dynamic humans were simulated using Doppler-enhanced array processing while the images of the static humans were generated from measurement data using range-enhanced array processing. Before denoising, the images were considerably distorted by through-wall propagation effects. Our algorithm showed that after denoising, the images were structurally similar with low mean square error with respect to the free space images. 
\section*{Acknowledgement}
This work is sponsored by the Ministry of Electronics and Information Technology, Govt. of India,under the Visvesvaraya PhD Scheme and 5IOA036 FA23861610004 grant by Air Force Office of Scientific Research (AFOSR), AOARD.
\appendix
\subsection{Hyper-parameter Selection}
\label{appendixxa}
\begin{figure*}[htbp]
\centering 
\includegraphics[scale=0.60]{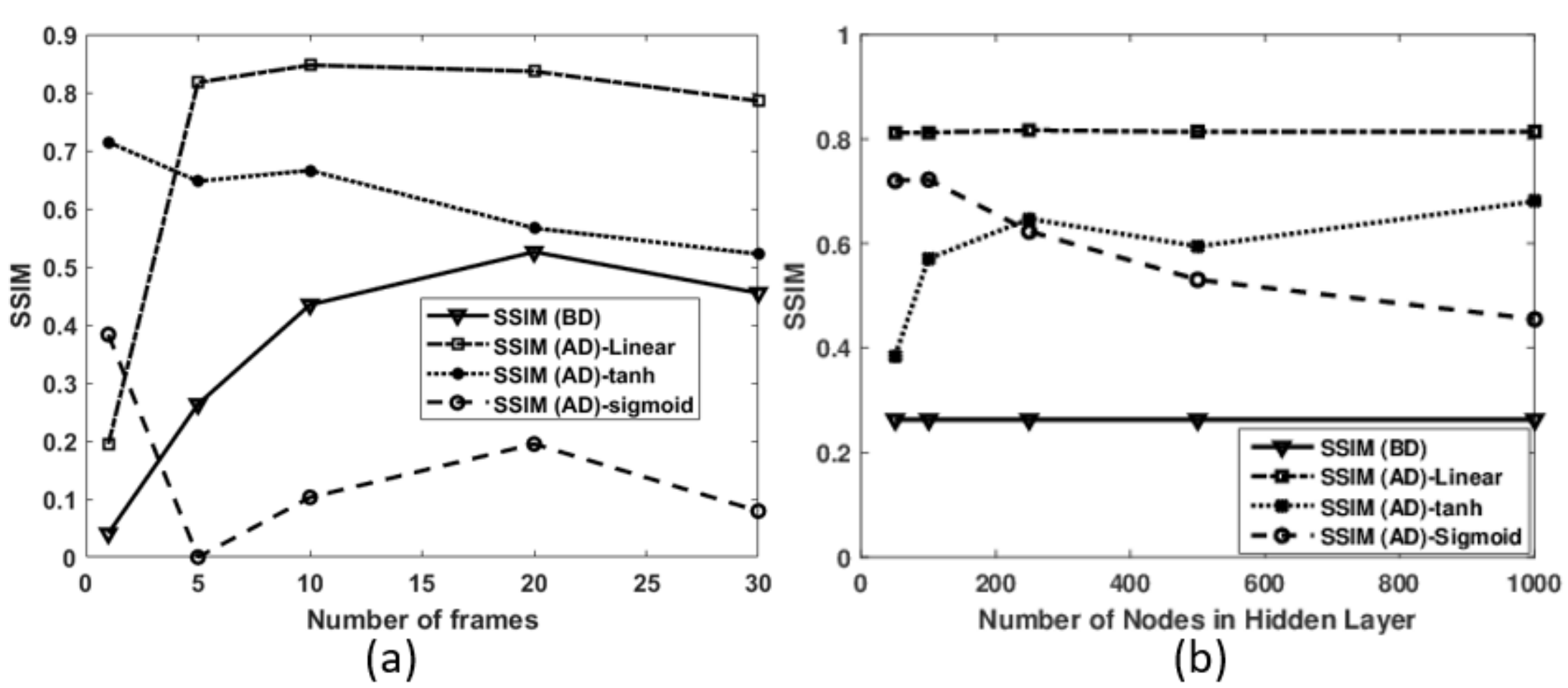}
\hfil \caption{SSIM variation for simulation results with respect to (a) number of frames, (b) number of nodes in the hidden layer for mapping functions-linear, tanh and sigmoid}
\label{fig:simulation_results_extra}
\end{figure*}

\begin{figure}
\centering
\includegraphics[scale=0.35]{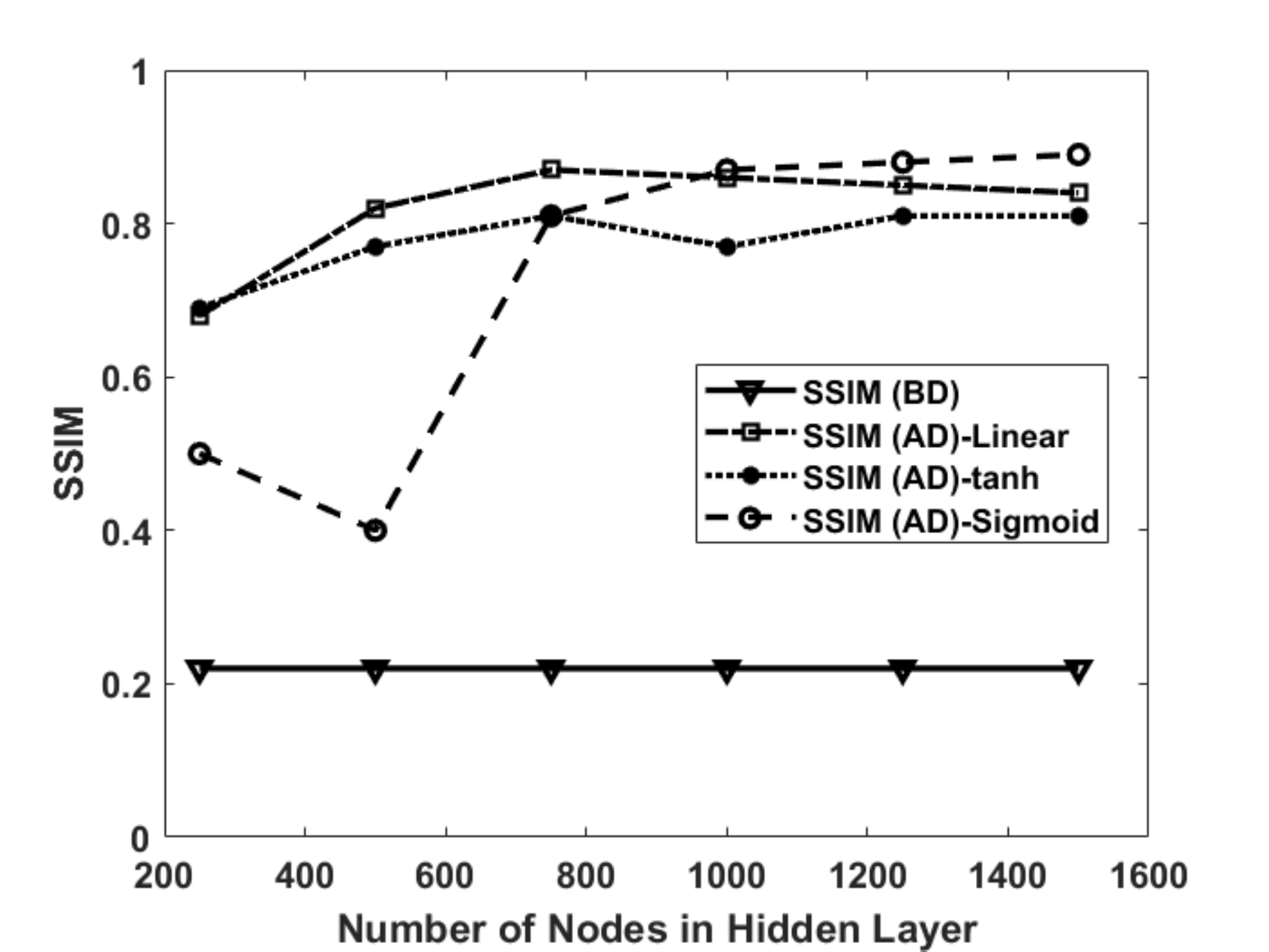}
\hfil
\caption{SSIM variation for measurement results with respect to number of nodes in the hidden layer for mapping functions-linear, tanh and sigmoid for human subjects}
\label{fig:Exp_variation_results}
\end{figure}

We optimized the autoencoder network's following parameters-number of nodes in the hidden layer and mapping functions to obtain the results presented in the previous sections. First we discuss the autoencoder used on the simulation data. Fig.\ref{fig:simulation_results_extra}  shows the performance for different mappings- linear, tanh, sigmoid as a function of number of frames in Fig.\ref{fig:simulation_results_extra}(a). We observe that the linear mapping outperforms the results obtained using non-linear mapping functions. We hypothesize that this is the case because the wall response remained mainly linear for narrowband measurements. Fig.\ref{fig:simulation_results_extra}(b), show the variation of SSIM before and after denoising as a function of the number of nodes in the hidden layer. We observe that the performance converges when the number of nodes is approximately 500. 
Next, we discuss the autoencoder used on the measurement data of real humans in both line-of-sight and through-wall conditions.The SSIM varies as a function of the number of nodes in the hidden layer for different mapping functions in Fig. \ref{fig:Exp_variation_results}. The SSIM improves and tends towards 1 as we increase the number of nodes in the hidden layer to 1500. The best performance is for the sigmoid non-linear function possibly because of the wideband nature of the propagation phenomenology. 
\bibliographystyle{IEEEtran}
\bibliography{arXiv}
\end{document}